\documentclass[pre,aps,onecolumn,superscriptaddress]{revtex4-1}
\usepackage{amsmath, amsthm, amssymb}
\usepackage{amsfonts}
\usepackage{graphicx}
\usepackage{dcolumn}
\usepackage{bm}
\usepackage{caption}
\usepackage{subcaption}
\usepackage{textcomp}

\usepackage{color}

\begin{document}

\title{Statistics of large currents in the Kipnis-Marchioro-Presutti model in a ring geometry}
\author{Lior Zarfaty}
\author{Baruch Meerson}
\email{meerson@mail.huji.ac.il}
\affiliation{Racah Institute of Physics, Hebrew University of
Jerusalem, Jerusalem 91904, Israel}

\begin{abstract}
\noindent
We use the macroscopic fluctuation theory to determine the statistics of large currents in the Kipnis-Marchioro-Presutti (KMP) model in a ring geometry. About 10 years ago this simple setting was instrumental
in identifying a breakdown of the additivity principle in a class of lattice gases at currents exceeding a critical value. Building on earlier work, we assume that, for supercritical currents, the optimal density profile, conditioned on the given current, has the form of a traveling wave (TW). For the KMP model we find this TW analytically, in terms of elliptic functions, for any supercritical current $I$. Using this TW solution, we evaluate, up to a pre-exponential factor, the probability distribution $P(I)$. We obtain simple asymptotics of the TW and of $P(I)$ for currents close to the critical current, and for currents much larger than the critical current. In the latter case we show that $-\ln P (I) \sim I\ln I$, whereas the optimal density profile acquires a soliton-like shape. Our analytic results are in a very good agreement with Monte-Carlo simulations and numerical solutions of Hurtado and Garrido (2011).
\end{abstract}

\maketitle

\tableofcontents
\nopagebreak

\section{Introduction}
\label{intro}

Diffusive lattice gases describe simple classical transport models \cite{spohn,L99,KL99,KRB10}. Besides their other applications, they have been extensively used for studying fluctuations of the density and current far from thermal equilibrium \cite{SZ95,S00,derrida2007,BE07,KRB10}. One of the simplest settings here is a one-dimensional ring which involves a large number of lattice sites \cite{derrida2007,Hurtado2014}. Because of the periodic boundary conditions, the average current through the ring vanishes. Fluctuating currents, however, are non-zero,  and it is interesting to find the probability distribution of observing a given current in a certain time interval. One simplifying hypothesis (both here and in other, non-ring, settings)  is known under the name of ``additivity principle". It assumes that the optimal density profile of the gas, conditional on a given current, is time-independent, leading to Gaussian statistics of the current \cite{derrida2004}. Indeed, for not too large currents, the additivity principle was
verified in Monte-Carlo simulations of the Kipnis-Marchioro-Presutti (KMP) model on an interval the boundaries of which are kept at different temperatures \cite{Hurtado2009}. This model was originally suggested as a microscopic model for which, at a coarse-grained level, the Fourier's law of heat conduction can be rigorously proven \cite{kipnis1982}. The model consists of a lattice of agents who carry a continuous amount of energy. At each stochastic move the energy is redistributed, via the uniform distribution, among a randomly chosen pair of nearest neighbors. This process conserves the energy locally, and will conserve it globally under appropriate boundary conditions, including those of a ring.

The additivity principle has also been found to hold in several other settings, all dealing with
large deviations of current in conservative and non-conservative lattice gases \cite{Hurtado2,MVK,M15,AMV}.
However, already in 2005 it was found that, for some lattice gases on a ring, the additivity principle breaks down when the current exceeds a critical value \cite{bertini2005,derrida2005}. When this happens, the system undergoes a dynamical phase transition, and the optimal density profile becomes time-dependent. One lattice gas model that exhibits breakdown of the additivity principle in the ring geometry is the KMP model, and the dynamical phase transition in this model was clearly identified in stochastic simulations  \cite{Hurtado2011}.

A coarse-grained description of diffusive lattice gases, including the KMP model, is provided by a Langevin equation \cite{spohn}. In one spatial dimension we have
\begin{equation}
    \label{continuity}
    \partial_t q + \partial_x j = 0,
\end{equation}
where
\begin{equation}
    \label{langevin}
    j = -D(q)\partial_x q + \sqrt{\sigma(q)}\eta.
\end{equation}
Here $q=q(x,t)$ is the energy density, $j=j(x,t)$ is the current density, and $\eta=\eta(x,t)$ is a delta-correlated Gaussian noise which satisfies
\begin{equation}
    \label{noise_term}
    \left<\eta(x,t)\right> = 0, \quad \left<\eta(x,t)\eta(x',t')\right> = \delta(x-x')\delta(t-t') .
\end{equation}
The gas diffusivity $D(q)$ and mobility $\sigma(q)$ are determined by the microscopic dynamics of the specific model. For the KMP model \cite{spohn}
\begin{eqnarray}
    \label{diffusivity} D(q)      &=& D_0,    \\
    \label{mobility}    \sigma(q) &=& 2aD_0q^2,
\end{eqnarray}
where  $D_0 =\text{const}$ and $a$ is the lattice constant. In the literature, a dimensionless description of the transport coefficients is often used, where $D_0=a=1$.

In this work we will use the macroscopic fluctuation theory (MFT): a large-deviation theory for the Langevin
equation (\ref{continuity}) and (\ref{langevin}).  The MFT was originally developed by Bertini \emph{et al}. for studying the non-equilibrium steady states of driven diffusive lattice gases, see Ref.~\cite{bertini2015} and references therein. The MFT is a weak-noise theory based on a saddle-point evaluation of the exact path integral of the Langevin equation. The MFT leads to a variational formulation for the optimal density profile, conditioned on a given large deviation. A closely related approach is the optimal fluctuation method that goes back to Refs. \cite{Halperin,Langer,Lifshitz}, see also Ref. \cite{LGP}.  Being especially suitable for sufficiently steep distribution tails, similar weak-noise theories have been applied to turbulence \cite{turb1,turb2,turb3}, stochastic reactions \cite{EK,MS2011}, non-equilibrium surface growth and related models \cite{Fogedby,KK,MKV,MV}, and other systems. The MFT equations can be formulated as a classical Hamiltonian field theory. Having solved the MFT equations, one can evaluate the action functional, from which the probability to observe a specific large deviation is obtained up to a sub-leading pre-factor.

Bodineau and Derrida \cite{derrida2005} showed that, for currents larger than the critical current, the optimal density profile, conditioned on the given current, corresponds to a traveling wave (TW) solution of the
MFT equations. They found an implicit TW solution, in the form of a first-order ordinary differential equation, with two integral constraints, for a general lattice gas on a one-dimensional ring  \cite{derrida2005}. Hurtado and Garrido \cite{Hurtado2011} performed Monte-Carlo simulations of the KMP model on a ring. As they observed,
``... for currents above a critical threshold the system self-organizes into a coherent traveling wave which
facilitates the current deviation by gathering energy in a localized packet, thus breaking translation
invariance" \cite{Hurtado2011}. Hurtado and Garrido \cite{Hurtado2011} also found the TW solution numerically. Meerson and Sasorov \cite{meerson2013} studied the large-current statistics of the KMP model on an infinite interval with a step-like density at $t=0$. They found that the optimal time history of a very large current fluctuation has the form of a (slowly evolving) soliton-like pulse, while the probability  of a large current obeys a sub-Gaussian statistics of the form $\ln P \propto -j\ln j$, where $j$ is a rescaled current.

Building on these works, here we study in detail the TW solution to the MFT equations for the KMP model on a ring:
for all currents, including arbitrary large ones. We find the TW solution and the resulting probability analytically. Our equation for the density profile coincides with the equation obtained by Bodineau and Derrida \cite{derrida2005}, and we solve it explicitly for the KMP model. We obtain simple asymptotics of the density profile and action close to the phase transition and in the limit of large currents. In the former limit, the results of Bodineau and Derrida are reproduced and extended to higher orders. In the latter limit, we observe a similar (but simpler) behavior of the solution compared with the one derived by Meerson and Sasorov \cite{meerson2013} for the infinitely long system. Finally, we compare our analytical results with the Monte-Carlo simulations and numerical solutions of Hurtado and Garrido \cite{Hurtado2011} for the KMP model, and observe a very good agreement.

The remainder of this paper is organized as follows. In Sec. \ref{MFT_sec} we present the governing MFT equations for the optimal density profile on a one-dimensional ring, and the boundary conditions and constraints. We reproduce the time-independent, constant-density solution which, for models such as the KMP, serves as the optimal profile only for subcritical currents. In Sec. \ref{scaling_sec} we discuss the general scaling behavior of the action, and the particular form of the scaling obtained under the TW assumption. In Sec. \ref{DODP} we derive the optimal density profile, conditional on a specific supercritical current, assuming a TW solution. In Sec. \ref{action_sec} we calculate, analytically and numerically, the action and other attributes of the TW solution. Sections \ref{mildly} and \ref{soliton} presents simple closed-form asymptotics:  close to the critical current and in the limit of very large currents, respectively. In Sec. \ref{comparison} we compare our analytic results with the Monte-Carlo simulations and numerical solutions of Hurtado and Garrido \cite{Hurtado2011} for the KMP model.  We summarize our work in Sec. \ref{disc}.

\section{The MFT formalism on a one-dimensional ring}
\label{MFT_sec}

\subsection{The MFT equations and constraints}
\label{MFT_general}

Equations (\ref{continuity})-(\ref{noise_term}) need to be supplemented by boundary conditions in space and in time, and by other system-dependent constraints. A ring of length $L$ enforces the periodic boundary conditions
\begin{equation}
    \label{boundary_cond}
    q(0,t) = q(L,t), \quad \eta(0,t) = \eta(L,t).
\end{equation}
A generic initial condition is of the form
\begin{equation}
    \label{initial_cond}
    q(x,0)=q_0(x), \quad q_0(0) = q_0(L).
\end{equation}
As we are interested in the current statistics, we need to constrain the realizations of $\eta(x,t)$ to those which produce a specific current
\begin{equation}
    \label{constraint_current}
    \frac{1}{LT} \int_0^T \!dt \int_0^L \!dx \,j = J,
\end{equation}
where $T$ is the measurement time. As the total energy is conserved, we can write
\begin{equation}
    \label{constraint_mass}
    \frac{1}{L} \int_0^L \!dx \,q = n_0,
\end{equation}
where $n_0$ is the average density of the gas. Equation~(\ref{constraint_mass}) will be necessary when
we look for a TW solution, which disregards the initial condition except for this constraint.

Using path integral formalism (see e.g. Ref.~\cite{path_integrals}), the probability density functional $P[\eta]$ of the noise term $\eta$ can be expressed as
\begin{equation}
    \label{probability_noise}
    \int \! D\eta \, P[\eta] = \int \! D\eta \, \exp \left( -\int_0^T \! dt \int_0^L \! dx \, \frac{\eta^2}{2} \right) = 1,
\end{equation}
from which the probability of a specific realization of $q$ can be obtained with the help of the Langevin equation (\ref{langevin}),
\begin{equation}
    \label{probability_density}
    -\ln\left(P[q]\right) \, \simeq \int_0^T \! dt \int_0^L \! dx \, \frac{\left[ j + D(q)\partial_x q \right]^2}{2\sigma(q)} = S[q],
\end{equation}
where $q$ and $j$ are coupled via the continuity equation (\ref{continuity}), and we have defined the action functional $S[q]$. The MFT equations can be derived from a saddle-point minimization  of the action functional, see Appendix \ref{derivation}. The resulting equations are
\begin{eqnarray}
    \label{MFT_equation_q}  \partial_t q &=& \partial_x \left[   D(q)\partial_x q - \sigma(q)v \right],  \\
    \label{MFT_equation_v}  \partial_t v &=& \partial_x \left[ - D(q)\partial_x v - \frac{1}{2}\sigma'(q)v^2 \right],
\end{eqnarray}
where $v=\partial_x p$ is the gradient of the conjugate momentum density. During the minimization process, a temporal boundary condition on $v$ appears:
\begin{equation}
    \label{current_constraint_MFT}
    v(x,T) = \frac{\lambda}{LTJ},
\end{equation}
where $\lambda$ is a dimensionless Lagrange multiplier used to enforce the integral constraint (\ref{constraint_current}). The current density $j$ can be expressed through $q$ and $v$:
\begin{equation}
    \label{current density}
    j = \sigma(q)v - D(q)\partial_x q.
\end{equation}
Once the optimal path is found, the probability of the current $J$ can be evaluated up to a pre-exponential factor:
\begin{equation}\label{Pgeneral}
   -\ln P(J)   \simeq S(J) = \frac{1}{2}\int_0^T dt \int_0^L dx\,\sigma(q) v^2 .
\end{equation}

\subsection{Constant-density solution}
\label{CDSL}

The time-independent constant-density solution of Eqs.~(\ref{MFT_equation_q}) and (\ref{MFT_equation_v}) has the form
\begin{equation}
    \label{constant_solution}
              q = n_0, \quad v = \frac{J}{\sigma(n_0)} .
\end{equation}
This leads to
$$
S(J) = \frac{TL}{2}\frac{J^2}{\sigma(n_0)},
$$
which describes a Gaussian distribution $P(J)$. As mentioned above, for some lattice gases this solution ceases to be the action minimizer when the current exceeds a critical value $J_c$.

\section{Scaling behavior of the action for the KMP model}
\label{scaling_sec}

Some interesting information can be extracted from dimensional analysis of the MFT equations for the KMP model. In particular, the dimensional analysis identifies a parameter (the rescaled current) which controls the phase transition and the asymptotics of the solution.

\subsection{General scaling behavior}
\label{KMP_gen_scale}

Using the KMP transport coefficients (\ref{diffusivity}) and (\ref{mobility}), we can rewrite the  MFT equations (\ref{MFT_equation_q}) and (\ref{MFT_equation_v}) as
\begin{eqnarray}
    \label{MFT_equation_q_KMP} \partial_t q &=& D_0 \partial_x \left( \partial_x q - 2a q^2 v \right), \\
    \label{MFT_equation_v_KMP} \partial_t v &=& D_0 \partial_x \left(-\partial_x v - 2a v^2 q \right).
\end{eqnarray}
The constraints are
\begin{eqnarray}
    \label{constraint_mass_KMP}     \frac{1}{L} \int_0^L \! dx \, q                        &=& n_0,   \\
    \label{constraint_current_KMP}  \frac{aD_0}{LT} \int_0^T \! dt \int_0^L \! dx \, 2q^2v &=& J.
\end{eqnarray}
The deterministic contribution to the current in Eq.~(\ref{constraint_current_KMP}) vanishes because of the periodic boundary conditions for $q$. The action is given by
\begin{equation}
    \label{action_MFT}
    S = aD_0 \int_0^T \! dt \int_0^L \! dx \, q^2v^2.
\end{equation}
Let us rescale these equations and constraints, without making any assumptions about the character of the solution. Upon the change of variables
\begin{equation}
    \label{scaling_variable_change}
              \bar{x} = \frac{x}{L} ,\quad \bar{t} = \frac{D_0t}{L^2} ,\quad \bar{q}(\bar{x},\bar{t}) = \frac{q(x,t)}{n_0} ,\quad \bar{v}(\bar{x},\bar{t}) = aLn_0v(x,t),
\end{equation}
the equations become
\begin{eqnarray}
    \label{MFT_equation_q_KMP_norm} \partial_{\bar{t}}\bar{q} &=& \partial_{\bar{x}} \left(  \partial_{\bar{x}}\bar{q} - 2\bar{q}^2\bar{v} \right), \\
    \label{MFT_equation_v_KMP_norm} \partial_{\bar{t}}\bar{v} &=& \partial_{\bar{x}} \left( -\partial_{\bar{x}}\bar{v} - 2\bar{v}^2\bar{q} \right).
\end{eqnarray}
The constraints are
\begin{eqnarray}
    \label{constraint_mass_KMP_norm}     \int_0^1 \! d\bar{x} \, \bar{q} &=& 1,  \\
    \label{constraint_current_KMP_norm}  \frac{1}{\bar{T}} \int_0^{\bar{T}} \! d\bar{t} \int_0^1 \! d\bar{x} \, 2\bar{q}^2\bar{v} &=& \frac{LJ}{D_0n_0} \equiv I,
\end{eqnarray}
where $I$ is the rescaled current, and $\bar{T}=D_0T/L^2$ is the rescaled measurement time. The action is given by
\begin{equation}
    \label{action_MFT_norm}
    S = \frac{L}{a} \int_0^{\bar{T}} \! d\bar{t} \int_0^1 \! d\bar{x} \, \bar{q}^2\bar{v}^2 \equiv \frac{L}{a} S_1\left(\bar{T},I\right).
\end{equation}
In the absence of a TW solution, $S_1(\bar{T},I)$ would be the large-deviation function of the current. Finally, the rescaled form of the constant-density solution, which appears in subsection \ref{CDSL}, is
\begin{eqnarray}
    \label{constant_solution_KMP}
    \bar{q} = 1, \quad \bar{v} = \frac{I}{2}, \quad S_1 = \frac{1}{4}\,\bar{T} I^2 .
\end{eqnarray}
We will suppress the bars in the following.

\subsection{Traveling wave scaling}
\label{TWsca}

From now on, we assume that the dominant contribution to the action, for a given supercritical current, comes from a TW solution
\begin{equation}
    \label{TW_assumption}
              q(x,t) = q(\xi), \quad v(x,t) = v(\xi), \quad \xi = x-ct,
\end{equation}
where $c$ is the (a priori unknown) speed of the TW. Using this ansatz in Eqs.~(\ref{constraint_mass_KMP_norm}) and (\ref{constraint_current_KMP_norm}), we obtain
\begin{eqnarray}
    \label{constraint_mass_TW}     \int_0^1 \! d\xi \, q       &=& 1,\\
    \label{constraint_current_TW}  \int_0^1 \! d\xi \, 2q^2v   &=& I.
\end{eqnarray}
For the action (\ref{action_MFT_norm}) we have
\begin{equation}
    \label{action_TW_tilde}
    S_1\left(\bar{T},I\right) = \bar{T} \int_0^1 \! d\xi \, q^2v^2 = \bar{T} s_1(I),
\end{equation}
therefore
\begin{equation}
    \label{action_TW}
  -\ln P\simeq  S = \frac{D_0T}{aL} s_1\left(\frac{LJ}{D_0n_0}\right) .
\end{equation}

\section{The TW solution}
\label{DODP}

In this and the following section, we find the exact TW solution of the MFT equations, and use it to derive an analytical expression for the action $s_1(I)$ from Eq.~(\ref{action_TW}). The TW solution cannot hold for all of the time $T$, because it obeys neither the temporal boundary condition (\ref{current_constraint_MFT}), nor the generic initial condition (\ref{initial_cond}). Essentially, we assume here that there are narrow  boundary layers in time: at the beginning and the end of the interval $0<t<T$, where the TW solution adapts to the boundary conditions in time. Similar narrow boundary layers in time are to be expected (or have been already observed numerically) in different settings where a simple solution of the MFT equations (a steady state or a TW)  dominates contribution to the action,  but does not satisfy one or both of the boundary conditions in time.

\begin{figure}[ht]
\includegraphics[scale=0.9]{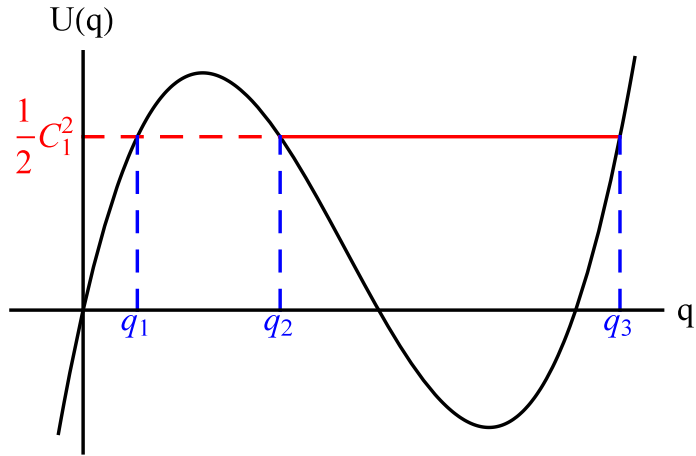}
\caption{The effective potential~(\ref{potential_def}). The $q_i$'s are the points for which $q'=0$. The motion is allowed in the region of $q_2<q<q_3$ (the solid straight line). The weakly supercritical regime (when the current is close to the critical current) occurs when $q_2$ approaches $q_3$. Here the motion is close to harmonic. In the strongly nonlinear regime (for very large currents) $q_2$ approaches $q_1$, and one observes a soliton-like density profile. If one sets $C_1<0$ or $C_2<0$, the possible range of $q$ will not allow a smooth transition between these two limits, due to the demand that $q\geq 0$ for all $\xi$.}
\label{potential}
\end{figure}

Plugging the ansatz (\ref{TW_assumption}) into (\ref{MFT_equation_q_KMP_norm}) and (\ref{MFT_equation_v_KMP_norm}), and performing integrations with respect to $\xi$ yields two first-order ordinary differential equations
\begin{eqnarray}
    \label{MFT_equation_q_TW}   q' &=& - cq + 2q^2v + C_1, \\
    \label{MFT_equation_v_TW}   v' &=&   cv - 2v^2q - C_2,
\end{eqnarray}
where the primes denote the derivative with respect to $\xi$, and $C_1$ and $C_2$ are yet unknown constants. Equations~(\ref{MFT_equation_q_TW}) and (\ref{MFT_equation_v_TW}) are Hamiltonian, with the Hamiltonian
\begin{equation}
    \label{conservation_law_TW}
    H = - cqv + q^2v^2 + C_1v + C_2q = E,
\end{equation}
a conserved quantity. Using Eq.~(\ref{conservation_law_TW}) in Eq.~(\ref{MFT_equation_q_TW}), one obtains a single first-order
equation \cite{derrida2005}
\begin{equation}
    \label{equation_for_q}
    q' = \pm \sqrt{ (C_1-cq)^2 - 4q^2(C_2q-E) }.
\end{equation}
In view of the obvious mechanical analogy, let us rewrite it as
\begin{equation}
    \label{potential_def0}
    \frac{1}{2}q'^2+U(q) = \frac{1}{2}C_1^2,
\end{equation}
where
\begin{equation}
    \label{potential_def}
   U(q) = 2C_2q^3 - \frac{1}{2}q^2(c^2+4E) + C_1cq
\end{equation}
is the effective potential. Reasonable solutions are obtained when $C_1$ and $C_2$ are positive. The potential $U(q)$ is depicted in Fig.~(\ref{potential}). Let us define $q_3>q_2>q_1>0$ as the values of $q$ for which $q'=0$. The effective particle motion is allowed in the region of $q_2<q<q_3$. (Back to the original problem,  $q_2$ and $q_3$ are the minimum and the maximum of the density profile, respectively.)  When  $q_2$ is close to $q_3$, the effective particle motion is close to harmonic. This is the  weakly supercritical regime, observed close to the critical current.  In the strongly nonlinear regime, corresponding to very large currents,  $q_1$ and $q_2$
become close to each other. The effective particle phase trajectory then approaches the homoclinic trajectory,
and the waveform approaches that of a soliton \cite{meerson2013}.

Rewriting Eq.~(\ref{equation_for_q}) in terms of $q_1$, $q_2$ and $q_3$, we obtain
\begin{equation}
    \label{equation_for_q_rewrite}
    q' = \pm 2\sqrt{ C_2(q-q_1)(q-q_2)(q_3-q) } .
\end{equation}
This equation can be integrated to yield the solution in terms of the Jacobi elliptic function $\text{dn}(u,k)$ \cite{elliptic_functions}:
\begin{equation}
    \label{q_sol_3}
    q(\xi) = q_1 + (q_3-q_1)\,\text{dn}^2 \left[\sqrt{C_2(q_3-q_1)}\,\xi , k \right],
\end{equation}
where
\begin{equation}
    \label{modulus_def}
    k = \sqrt{\frac{q_3-q_2}{q_3-q_1}}
\end{equation}
is the elliptic modulus. We have omitted in Eq.~(\ref{q_sol_3}) an arbitrary constant, resulting from translational symmetry of the solution, and thus have set
the density peak to be at $\xi=0$. Now we can find a closed expression for $v(\xi)$. After some algebra,
\begin{equation}
    \label{v_structure}
    v(\xi) = V\left\{ q_1 + (q_3-q_1) \,\text{dn}^2 \left[ \sqrt{C_2(q_3 - q_1)}(\xi + \phi) , k \right] \right\}.
\end{equation}
where
\begin{equation}
    \label{v_structure_sol}
    V = \frac{1}{2}\sqrt{\frac{C_2}{q_1q_2q_3}} ,\quad \phi = \frac{1}{ \sqrt{C_2(q_3-q_1)} }\,\text{arcdn} \left( \sqrt{\frac{q_2}{q_3}} , k \right) ,
\end{equation}
where $\text{arcdn}(u,k)$ is one of the inverse Jacobi elliptic functions \cite{elliptic_functions}.  The solution includes four constants: $C_2$, $q_1$, $q_2$ and $q_3$ that need to be found.  Integrating Eq~(\ref{MFT_equation_q_TW}) with respect to $\xi$ and using the constraints (\ref{constraint_mass_TW}) and (\ref{constraint_current_TW}), we obtain $I=c-C_1$ and
\begin{equation}
    \label{C2_exact_sol}
    \sqrt{C_2} = \frac{I}{2\sqrt{q_1q_2q_3}}\left(\frac{q_1q_2+q_2q_3+q_3q_1}{2q_1q_2q_3}-1\right)^{-1}.
\end{equation}
Because of the periodic boundary conditions, the system length (which, in the rescaled units, is 1) must contain an integer number of periods of the oscillating function $q(\xi)$. The minimum action, however,
is achieved for the ``fundamental mode" (see Appendix \ref{multiple_waves}), so we demand
\begin{equation}
    \label{constraint_qs1}
    \frac{2\text{K}(k)}{\sqrt{C_2(q_3-q_1)}} = 1,
\end{equation}
where $\text{K}(k)$ is the complete elliptic integral of the first kind \cite{elliptic_integrals}.
Finally, the energy conservation (\ref{constraint_mass_TW}) yields
\begin{equation}
    \label{constraint_qs2}
    q_1+(q_3-q_1)\frac{\text{E}(k)}{\text{K}(k)} = 1,
\end{equation}
where $\text{E}(k)$ is the complete elliptic integral of the second kind \cite{elliptic_integrals}. Now we can rewrite the TW solution as
\begin{eqnarray}
   q(\xi) &=& q_1 + (q_3-q_1)\,\text{dn}^2 \left[ 2\text{K}(k)\xi , k \right], \nonumber \\
   v(\xi) &=& \frac{\text{K}(k)}{\sqrt{q_1q_2q_3(q_3-q_1)}} \left\{ q_1 + (q_3-q_1)\,\text{dn}^2 \left[ 2\text{K}(k)(\xi + \phi) , k \right] \right\},   \quad
    \phi =\frac{1}{2\text{K}(k)}\,\text{arcdn} \left( \sqrt{\frac{q_2}{q_3}} , k \right).
    \label{solution_density_final}
\end{eqnarray}
The solution includes three constants $q_1$, $q_2$ and $q_3$, and we have only two constraints: Eq.~(\ref{constraint_qs2}) and the equation
\begin{equation}
    \label{constraints10}
  4\text{K}(k)\sqrt{\frac{q_1q_2q_3}{q_3-q_1}}\left(\frac{q_1q_2+q_2q_3+q_3q_1}{2q_1q_2q_3}-1\right) = I,
\end{equation}
which results from Eqs.~(\ref{C2_exact_sol}) and (\ref{constraint_qs1}). Therefore, we will have to minimize the resulting action with respect to the last remaining constant \cite{derrida2005,Hurtado2011}.

\section{The TW action}
\label{action_sec}

Integrating equation (\ref{conservation_law_TW}) over $\xi$, we obtain
\begin{equation}
    \label{action_analytical_1}
    s_1 = E - 2C_2 + c\int_0^1 \!d\xi \, qv,
\end{equation}
where the rescaled action $s_1$ was defined in Eq.~(\ref{action_TW_tilde}). To evaluate the integral in Eq.~(\ref{action_analytical_1}), we can use the solution (\ref{solution_density_final}) for $q$ and $v$. The last term of the product $qv$ can be simplified using the identity \cite{elliptic_functions} (where we suppress the elliptic modulus $k$):
\begin{align}
    \label{jacobi_dn_quartic}
    \text{dn}^2\, u \,\text{dn}^2(u+a) = & 2\,\text{ds}\, a \,\text{ns}\,a\,\text{cs}\,a\,\left\{ \text{Z}\left[\text{am}(u+a)\right] -\text{Z}\left(\text{am}\,u\right) -\text{Z}\left(\text{am}\,a\right) \right\} \\ \nonumber
    & - \text{cs}^2\,a\left[ \text{dn}^2(u+a) + \text{dn}^2\,u\right] + \text{ds}^2\,a + \text{cs}^2\,a .
\end{align}
Here $\text{am}(u,k)$, $\text{Z}(u,k)$, $\text{ds}(u,k)$, $\text{ns}(u,k)$ and $\text{cs}(u,k)$ are the Jacobi amplitude function, the Jacobi zeta function and three of the Jacobi's elliptic functions, respectively \cite{elliptic_functions,elliptic_integrals}. Substituting $u=2K(k)\xi$ and $a=2K(k)\phi$ into Eq.~(\ref{jacobi_dn_quartic}) and using the periodicity of $\text{Z}\left[\text{am}(u,k),k\right]$, we obtain
\begin{align}
    \label{action_analytical_2}
    s_1 = \text{K}^2(k)\left\{ \frac{1}{q_3-q_1}\left[ \frac{q_2q_3}{q_1} + 2(q_2+3q_3-4) + q_1\left(\frac{q_3}{q_2} - \frac{3q_2}{q_3} + 2\right) \right] \vphantom{\frac{4(q_1q_2+q_2q_3+q_3q_1)}{\sqrt{q_1q_2q_3(q_3-q_1)}}} \right. \\ \nonumber
    \left. - \frac{4(q_1q_2+q_2q_3+q_3q_1)}{\sqrt{q_1q_2q_3(q_3-q_1)}} \text{Z} \left[ \arcsin\left(\sqrt{1-\frac{q_1}{q_3}}\right) , k \right] \right\}.
\end{align}
Now we need to minimizing the rescaled action (\ref{action_analytical_2}) subject to constraints~(\ref{constraint_qs2}) and (\ref{constraints10}). Introducing new constants
\begin{equation}
    \label{notation_switch_2}
    \gamma_2 = \frac{q_1}{q_2} ,\quad \gamma_3 = 1-\frac{q_1}{q_3},
\end{equation}
we can merge the two constraints (\ref{constraint_qs2}) and (\ref{constraints10}) into a single constraint on $\gamma_2$ and $\gamma_3$:
\begin{equation}
    \label{constraint_final}
    \frac{4\text{K}(k)}{\sqrt{\gamma_2\gamma_3}}\left\{ \frac{2+\gamma_2-\gamma_3}{2} - \left[ 1 + \frac{\gamma_3}{1-\gamma_3}\frac{\text{E}(k)}{\text{K}(k)} \right]^{-1} \right\} = I.
\end{equation}
The action becomes
\begin{equation}
    \label{action_analytical_final}
    s_1 = \text{K}^2(k)\left\{ 6 + \frac{4-3\gamma_3}{\gamma_2} + \frac{\gamma_2}{\gamma_3} -\frac{8\text{E}(k)}{\text{K}(k)} - \frac{4(2+\gamma_2-\gamma_3)}{\sqrt{\gamma_2\gamma_3}} Z \left[ \arcsin\left(\sqrt{\gamma_3}\right) , k \right] \right\},
\end{equation}
while the wave velocity and the phase shift can be written as
\begin{equation}
    \label{wave_phase_final}
    c = 2\text{K}(k) \frac{2+\gamma_2-\gamma_3}{\sqrt{\gamma_2\gamma_3}}, \quad \phi = \frac{1}{2\text{K}(k)}\text{arcdn} \left( \sqrt{\frac{1-\gamma_3}{\gamma_2}} , k \right).
\end{equation}
Minimizing $s_1$ with respect to $\gamma_2$, we obtain
\begin{equation}
    \label{constraint_final_2}
    \frac{ds_1}{d\gamma_2} = \frac{\partial s_1}{\partial \gamma_2} + \frac{\partial s_1}{\partial \gamma_3}\frac{d\gamma_3}{d\gamma_2} = 0,
\end{equation}
where $d\gamma_3/d\gamma_2$ can be obtained by differentiating the constraint (\ref{constraint_final}). As $q_3>q_2>q_1>0$, the constants $\gamma_2,\gamma_3 \in [0,1]$.

Let us first calculate the value of the critical current which we denote as $I_c$. The critical current corresponds to  $q_2 \rightarrow q_3$, see Fig.~(\ref{potential}). Therefore, $k \rightarrow 0$, and $\gamma_2+\gamma_3=1$. Plugging this relation into the action (\ref{action_analytical_final}) yields $s_1=\frac{\pi^2}{4\gamma_2(1-\gamma_2)}$. Minimizing this with respect to $\gamma_2$ on the interval $0<\gamma_2<1$, we obtain $\gamma_2=\gamma_3=1/2$. Then Eq.~(\ref{constraint_final}) yields $I_c=2\pi$, in agreement with Ref. \cite{derrida2005}. Now we realize that  $\gamma_2,\gamma_3 \in [\frac{1}{2},1]$, where the lower boundary corresponds to the critical current. In the large-current regime  $\gamma_2$ and $\gamma_3$ approach $1$.

Given a supercritical current $I>I_c$, one needs to solve the algebraic equations (\ref{constraint_final}) and (\ref{constraint_final_2}) for the constants $\gamma_2$ and $\gamma_3$. In Sections \ref{mildly} and \ref{soliton} we will find the corresponding asymptotics for the small supercriticality and for very large currents.  For intermediate currents we need to resort to numerics. For a given current, we find the constants $\gamma_2$ and $\gamma_3$ by a numerical minimization of the action (\ref{action_analytical_final}) subject to the constraint (\ref{constraint_final}). We then determine the wave velocity and phase shift using Eq.~(\ref{wave_phase_final}), as well as $q(\xi)$ and $v(\xi)$.

\begin{figure} [ht]
    \includegraphics[scale=1.0]{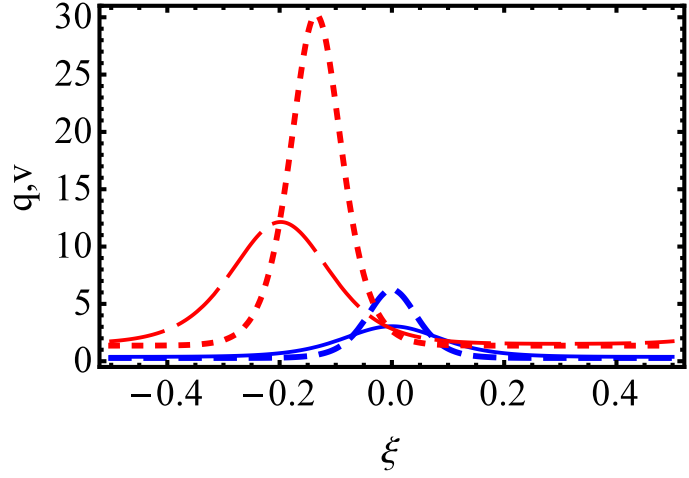}
    \caption{The TW solution~(\ref{solution_density_final}) for two values of the rescaled current $I$. Shown are the density field $q(\xi)$ and the canonically conjugate field $v(\xi)$ for currents $I=2I_c$ (solid and long-dashed, respectively) and $I=4I_c$ (medium-dashed and short-dashed, respectively). The field $q(\xi)$ travels ahead of $v(\xi)$; the two get closer when the current goes up. For $I \gg I_c$, the solution becomes soliton-like, see Sec. \ref{soliton}.}
    \label{solution_all}
\end{figure}

\begin{figure} [ht]
    \centering
    \includegraphics[scale=1.0]{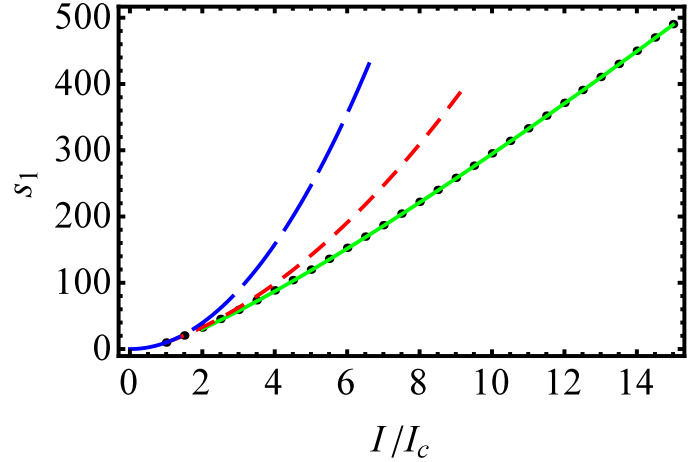}
    \caption{The rescaled action $s_1$, see Eqs.~(\ref{action_TW_tilde}) and~(\ref{action_TW}), versus the rescaled current. Shown are the exact TW action (symbols), the weakly supercritical TW asymptotic (medium-dashed), the large-current TW asymptotic (solid), and the constant-density action (long-dashed). The dynamic phase transition at $I=I_c$ is of the second order.  Surprisingly, the large-current approximation is quite accurate even for relatively small supercritical currents.}
    \label{action_all}
\end{figure}

\begin{figure} [ht]
    \centering
    \begin{subfigure}{0.45\textwidth}
        \caption{}
        \includegraphics[width=\textwidth]{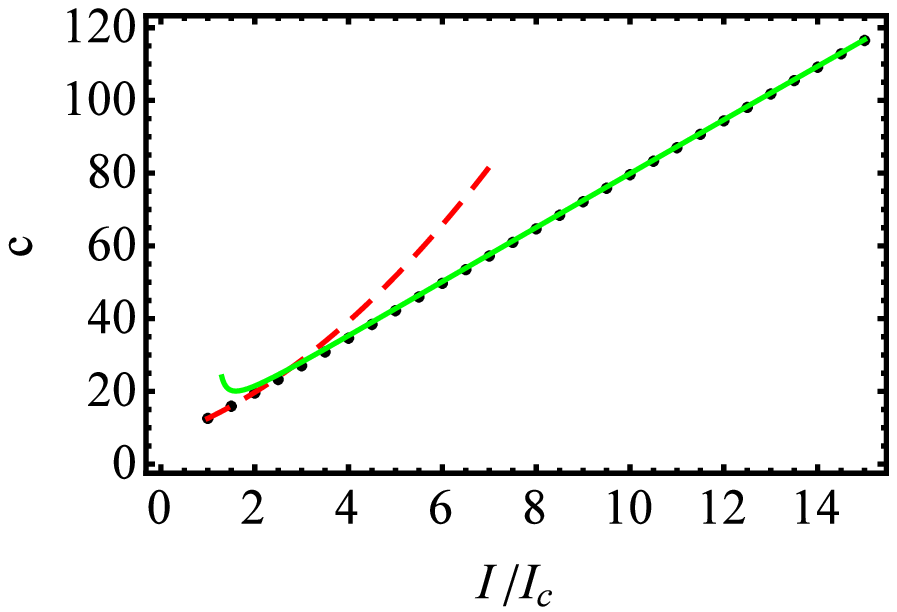}
        \label{velocity_all}
    \end{subfigure}
    \quad\quad
    \begin{subfigure}{0.45\textwidth}
        \caption{}
        \includegraphics[width=\textwidth]{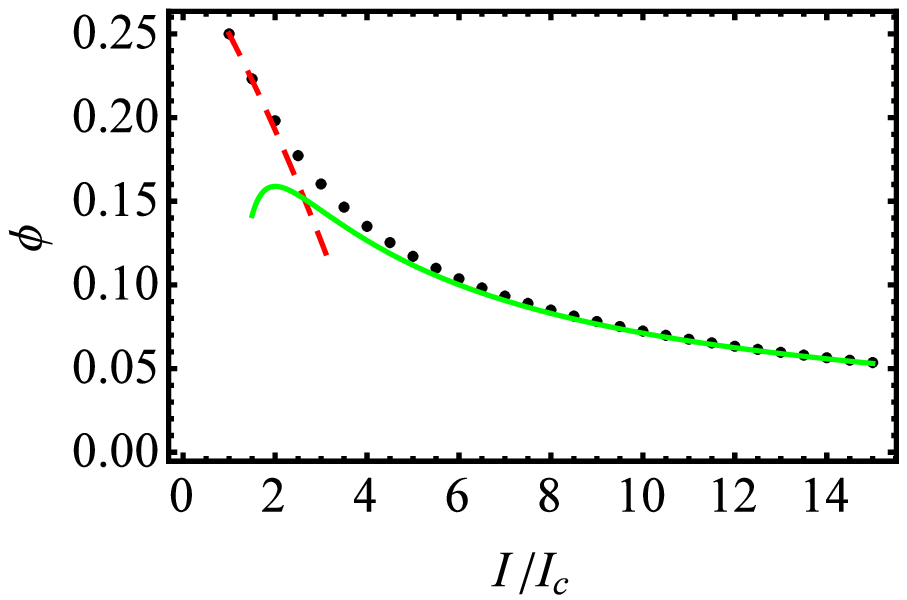}
        \label{phase_all}
    \end{subfigure}
    \caption{The TW velocity (a) and the phase shift between $q(\xi)$ and $v(\xi)$ (b) versus  the rescaled current. Shown are the exact results (symbols), the weakly supercritical asymptotic (dashed), and the large-current asymptotic (solid).}
    \label{velocity_phase_all}
\end{figure}

We observed that, as $I$ grows, $\gamma_2$ and $\gamma_3$ approach unity: $\gamma_2$ exponentially, $\gamma_3$ algebraically. As a result, a straightforward minimization of the action on the square $\gamma_2,\gamma_3 \in [1/2,1]$ is very difficult, because the sought values of $\gamma_1$ and $\gamma_2$ lie very close to the edges of the square. We worked around this numerical problem by introducing a new parametrization
\begin{equation}
    \label{numerics_param}
    \gamma_2 = 1 - \frac{1}{2\sqrt{1+d_2^2}} ,\quad \gamma_3 = 1 - \frac{1}{2\sqrt{1+d_3^2}},
\end{equation}
which maps the square of $[1/2,1] \times [1/2,1]$ to the whole real plane. Figure~\ref{solution_all} shows the resulting TW solution for two values of the supercritical current $I$. Figures~\ref{action_all}, \ref{velocity_all} and \ref{phase_all} shows the rescaled action $s_1$, the TW velocity $c$ and phase shift $\phi$, respectively, versus $I/I_c$, along with their asymptotics derived in the next two sections. The rescaled action
experiences a jump in the second derivative with respect to the current, implying a second-order dynamical phase transition for the large deviations of current \cite{derrida2005,Hurtado2011}.

An additional lattice gas, where the optimal profile, conditioned on a supercritical current, is believed to have the form of a TW,
is the weakly asymmetric exclusion process: WASEP \cite{derrida2005}. In that case Espigares et al. \cite{Espigares} obtained an exact solution for the TW profile in terms of elliptic functions, using a procedure similar to ours. The action was computed
in Ref. \cite{Espigares} numerically. As we showed here, for the KMP model it can be determined analytically. We also extracted asymptotics close to the critical current and at very large currents that we will now present.

\section{Close to the critical current}
\label{mildly}

When the current exceeds $I_c$, a lower-action TW solution bifurcates from the constant-density solution \cite{derrida2005}. Here we calculate the weakly-subcritical asymptotics for the rescaled action, TW velocity and phase shift, and find the shape of the solution in this limit. This is done by expanding all of the quantities and constraints in Taylor-like series with respect to the small parameter
\begin{equation}
    \label{small_parameter}
    \delta I = \frac{I}{I_c} - 1,
\end{equation}
or its positive powers. To identify the power of $\delta I$ in the expansions of $\gamma_2$ and $\gamma_3$,
we set $\gamma_2=1/2+\epsilon_2$, $\gamma_3=1/2+\epsilon_3$ and $I=2\pi(1+\delta I)$, and expand the constraint (\ref{constraint_final}) to lowest order in $\epsilon_2$, $\epsilon_3$ and $\delta I$. We obtain
\begin{equation}
    \label{first_order_scaling}
    -2\delta I + \frac{5\epsilon_2^2}{2} + \frac{5\epsilon_3^2}{2} + \epsilon_2\epsilon_3 = 0.
\end{equation}
Clearly, the scaling is $\epsilon_2 \sim \epsilon_3 \sim \sqrt{\delta I}$. Therefore, we set
\begin{eqnarray}
    \label{series_expansions_def}
    \gamma_2 & = & \frac{1}{2} + a_1\delta I^{1/2} + a_2\delta I + a_3\delta I^{3/2} + a_4\delta I^2 + ... \, , \\ \nonumber
    \gamma_3 & = & \frac{1}{2} + b_1\delta I^{1/2} + b_2\delta I + b_3\delta I^{3/2} + b_4\delta I^2 + ... \, .
\end{eqnarray}
One can then determine the series coefficients by substituting (\ref{series_expansions_def}) to the constraints (\ref{constraint_final}) and (\ref{constraint_final_2}), expanding in $\delta I$, and demanding that the constraints will hold for each order of the expansion, thus obtaining two sets of equations for the coefficients. This straightforward way, however, presents a difficulty, as the constraint (\ref{constraint_final_2}) turns out to be trivially satisfied for low orders of $\delta I$, demanding cumbersome high-order calculations. Instead, one can get one set of equations from the first constraint, then expand the action with respect to $\delta I$, and perform the minimization order by order. Using this approach, we obtained
\begin{eqnarray}
    \label{series_expansions_sol}
    \gamma_2 & = & \frac{1}{2} + \frac{1}{\sqrt{3}}\delta I^{1/2} + \frac{1}{6}\delta I - \frac{5}{24\sqrt{3}}\delta I^{3/2} - \frac{11}{72}\delta I^2 + \cdots, \\ \nonumber
    \gamma_3 & = & \frac{1}{2} + \frac{1}{\sqrt{3}}\delta I^{1/2} - \frac{1}{6}\delta I - \frac{5}{24\sqrt{3}}\delta I^{3/2} + \frac{11}{72}\delta I^2 +\cdots .
\end{eqnarray}
Now we can find the weakly-supercritical asymptotics for the wave velocity and phase shift, and the shape of the solution. The action has been already obtained as a part of the procedure of finding the expansions for $\gamma_2$ and $\gamma_3$. The results are
\begin{equation}
    \label{action_expansion}
    s_1 =\pi^2 \left( 1 + 2\delta I + \frac{1}{3}\delta I^2 +\cdots\right),
\end{equation}
\begin{equation}
    \label{velocity_phase_expansion}
    c = 4\pi \left( 1 + \frac{1}{2}\delta I + \frac{5}{72}\delta I^2 +\cdots \right) ,\quad \phi = \frac{1}{4} \left( 1 - \frac{2}{3\pi}\delta I - \frac{1}{18\pi}\delta I^2 +\cdots \right).
\end{equation}
These asymptotics are shown  in Figs.~\ref{action_all}, \ref{velocity_all} and \ref{phase_all}. As $I$ exceeds $I_c$, the TW starts off with a non-zero velocity. The period of the TW motion along the ring, back in the physical variables,
\begin{equation}
    \label{period_expansion}
    \tau = \frac{1}{c}\frac{L^2}{D_0} \simeq \frac{1}{4\pi} \left( 1 - \frac{1}{2}\delta I + \frac{13}{72}\delta I^2 \right) \frac{L^2}{D_0},
\end{equation}
is of the order of the diffusion time $L^2/D_0$. The density field $q(\xi)$ and the canonically conjugate field $v(\xi)$ are composed of harmonic waves, with the amplitude of the fundamental mode of order $\sqrt{\delta I}$:
\begin{eqnarray}
    \label{solution_density_expansion}
    q (\xi)& = & 1 + \frac{2}{\sqrt{3}}\cos(2\pi\xi)\,\delta I^{1/2} + \frac{2}{3}\cos(4\pi\xi)\,\delta I + \cdots, \\ \nonumber
    v (\xi)& = & \pi \left\{ 1 + \frac{2}{\sqrt{3}}\cos\left[2\pi(\xi+\phi)\right]\delta I^{1/2} +  \left[\frac{2}{3}\, \cos\left[4\pi(\xi+\phi)\right] + \frac{1}{3} \right] \delta I +\cdots \right\}.
\end{eqnarray}
\begin{figure} [ht]
    \centering
    \includegraphics[scale=1.0]{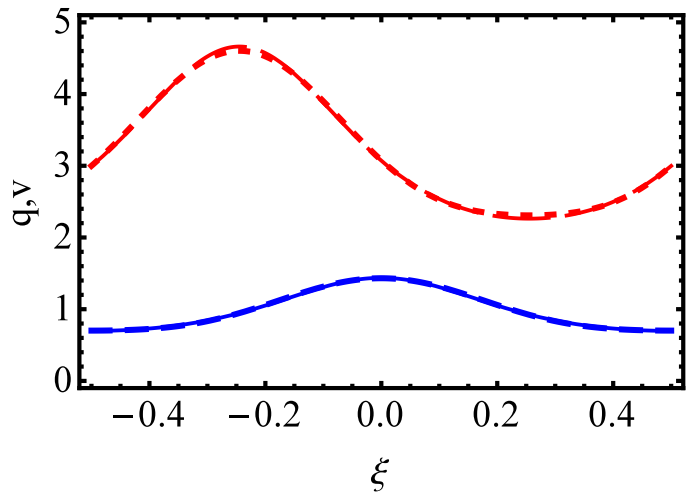}
    \caption{Exact solutions and asymptotics of the TW density field $q(\xi)$ and the canonically conjugate field $v(\xi)$ in the weakly supercritical regime for $\delta I = 0.1$. The solid and long-dashed lines represent the exact solutions (\ref{solution_density_final}) of $q$ and $v$, respectively. The medium-dashed and short-dashed lines represent the asymptotics (\ref{solution_density_expansion}) of $q$ and $v$, respectively.}
    \label{solution_small}
\end{figure}
Figure~\ref{solution_small} presents the exact solutions (\ref{solution_density_final}) for $q(\xi)$ and $v(\xi)$  and the weakly-supercritical asymptotics for $\delta I=0.1$.

Our weakly-supercritical results agree with the results of Bodineau and Derrida \cite{derrida2005} for the zero-order values of $s_1$, $c$ and $\phi$, the critical current $I_c=2\pi$, and the shape of the fundamental model $q \propto \cos(2\pi \xi)$. Here we have extended their zero-order results to the first and second orders in $\delta I$.

\section{Very large currents: the soliton}
\label{soliton}

We now consider very large currents, $I \gg I_c =\mathcal{O}(1)$ and calculate the asymptotics of the rescaled action, the TW velocity and the phase shift, and also find $q(\xi)$ and $v(\xi)$ in this limit. This is done by expanding all of the quantities and constraints with respect to the small parameter  $1/I$.

\subsection{Large-current expansion}

When the current is very large, $q_1$ approaches $q_2$, $q_3$ increases, and the TW acquires a soliton-like shape \cite{meerson2013}. In this limit both $\gamma_2$ and $\gamma_3$ approach 1, but in a different way. Let us introduce two small parameters,
\begin{equation}
    \label{ld_epsilon_def}
    \epsilon_2 = 1 - \gamma_2 ,\quad\epsilon_3 = 1- \gamma_3.
\end{equation}
Expanding Eqs.~(\ref{constraint_final}) and~(\ref{action_analytical_final}), one obtains a variety of terms depending differently on $\epsilon_2$ and $\epsilon_3$. Motivated by our numerical calculations, we assume the following scalings at $I\gg I_c=\mathcal{O}(1)$:
\begin{equation}
    \label{ld_epsilon_assumptions}
    \ln\left(\frac{1}{\epsilon_2\epsilon_3}\right) \sim I, \quad \mbox{and}\quad  \frac{1}{\epsilon_3} \sim I .
\end{equation}
(We are not assuming yet that $\ln I \gg 1$.) It is now possible to keep track of the different terms and determine which terms one needs to keep.

\subsection{Minimizing the action}

Keeping only leading order terms in the action (\ref{action_analytical_final}), we obtain
\begin{equation}
    \label{ld_action_temp}
    s_1 \simeq 2 \ln \left( \frac{16}{\epsilon_2\epsilon_3} \right) \ln \left( \frac{4}{e^2\epsilon_3} \right),
\end{equation}
while the constraint (\ref{constraint_final}) yields
\begin{equation}
    \label{ld_constraint_temp}
    \ln \left( \frac{16}{\epsilon_2\epsilon_3} \right) \simeq \frac{2I}{4-I\epsilon_3}.
\end{equation}
Plugging Eq.~(\ref{ld_constraint_temp}) into Eq.~(\ref{ld_action_temp}), we obtain $s_1$ as a function of $\epsilon_3$. Minimizing the action with respect to $\epsilon_3$ yields
\begin{equation}
    \label{epsilon3_sol}
    \epsilon_3 \simeq \frac{4}{I}\left|\text{W}_{-1}\left( -\frac{e}{I} \right)\right|^{-1},
\end{equation}
where $\text{W}_{-1}(u)$ is the secondary branch of the Lambert $W$-function \cite{lambertW}, which is defined for $u \in \left[\left.-1/e,0\right)\right.$ and has the following asymptotic expansion as $u \rightarrow 0^{-}$:
\begin{equation}
    \label{lambert_W_series}
    \left|\text{W}_{-1}(u)\right| = L_1 + L_2 + \frac{L_1}{L_2} + O\left[\left(\frac{L_1}{L_2}\right)^2\right], \quad L_1 = \ln\left(\frac{1}{|u|}\right), \quad L_2 = \ln\ln\left(\frac{1}{|u|}\right).
\end{equation}
This result is consistent with our scaling assumptions in (\ref{ld_epsilon_assumptions}).

\subsection{Leading-order behavior for $I \gg I_c$}

\begin{figure}
    \centering
    \includegraphics[scale=1.0]{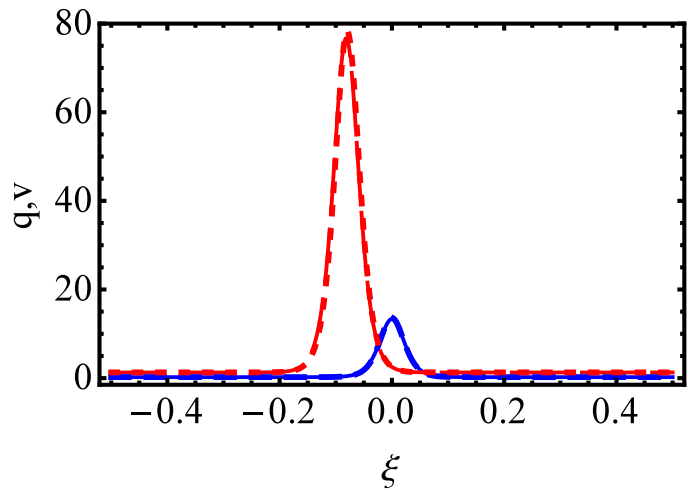}
    \caption{The exact TW density field $q(\xi)$ and the canonically conjugate field $v(\xi)$ alongside with their large-current soliton asymptotics for $I=8.5$. Solid and long-dashed: exact solution (\ref{solution_density_final}) of $q$ and $v$, respectively. Medium-dashed and short-dashed: asymptotics (\ref{ld_density_expansion}) of $q$ and $v$, respectively.}
    \label{solution_large}
\end{figure}

Using the identity
\begin{equation}\label{identity}
\text{W}_{-1}(u) = \ln \left[ \frac{u}{\text{W}_{-1}(u)} \right] ,
\end{equation}
we can rewrite $s_1$ in the following form:
\begin{equation}
    \label{ld_action_expansion}
    s_1 \simeq I \Omega, \quad \Omega = \left| \text{W}_{-1}\left( -\frac{e}{I} \right) \right|.
\end{equation}
Now we can find the large-current asymptotics of the wave velocity, phase shift, and the density profile. Expanding everything in the small parameters
(\ref{ld_epsilon_def}), using Eq.~(\ref{identity}) and keeping only leading order terms, we obtain
\begin{equation}
    \label{ld_velocity_phase_expansion}
    c \simeq \frac{I\Omega}{\Omega-1} ,\quad \phi \simeq \frac{\Omega^2-1}{I\Omega}.
\end{equation}
The predictions of Eqs.~(\ref{ld_action_expansion}) and (\ref{ld_velocity_phase_expansion}) are shown in Figs.~\ref{action_all}, \ref{velocity_all}, and \ref{phase_all}. The TW solution acquire a soliton-like shape
\begin{equation}
    \label{ld_density_expansion}
    q(\xi) \simeq \frac{I}{4} \,\text{sech}^2 \left( \frac{I\Omega}{\Omega-1} \frac{\xi}{2} \right) +\frac{1}{\Omega}, \quad
    v(\xi) \simeq \frac{\Omega}{\Omega-1} \left[  \frac{I\Omega}{4}\, \text{sech}^2 \left( \frac{I\Omega}{\Omega-1} \frac{\xi + \phi}{2} \right) +1\right].
\end{equation}
Figure~\ref{solution_large} depicts the exact solution (\ref{solution_density_final}), alongside with the soliton asymptotics (\ref{ld_density_expansion}) of $q(\xi)$ and $v(\xi)$, for $I=8.5 I_c$.

Let us now assume that the current is so large that $\ln I \gg 1$. Here $\Omega \simeq \ln I$, and we obtain
\begin{equation}
    \label{ld_extreme}
    s_1 \simeq I \ln I ,\quad c \simeq I ,\quad \phi \simeq \frac{\ln I}{I},
\end{equation}
whereas the TW solution simplifies to
\begin{equation}
    \label{ld_density_extreme_old}
    q(\xi) \simeq \frac{I}{4}\, \text{sech}^2 \left( \frac{I\xi}{2} \right), \quad
    v(\xi) \simeq \frac{I}{4}\, \ln I \,\text{sech}^2 \left[ \frac{I}{2}(\xi + \phi) \right].
\end{equation}
In this case, the period of the TW motion along the ring, in the physical units, is
\begin{equation}
    \label{period_ld_extreme}
    \tau = \frac{L^2}{cD_0} \simeq \frac{L^2}{I D_0},
\end{equation}
much smaller than the diffusion time $L^2/D_0$.

\section{Comparison to Monte-Carlo simulations and numerical solution of Hurtado and Garrido \cite{Hurtado2011}}
\label{comparison}

\begin{figure}
    \centering
    \begin{subfigure}{0.45\textwidth}
        \caption{}
        \includegraphics[width=\textwidth]{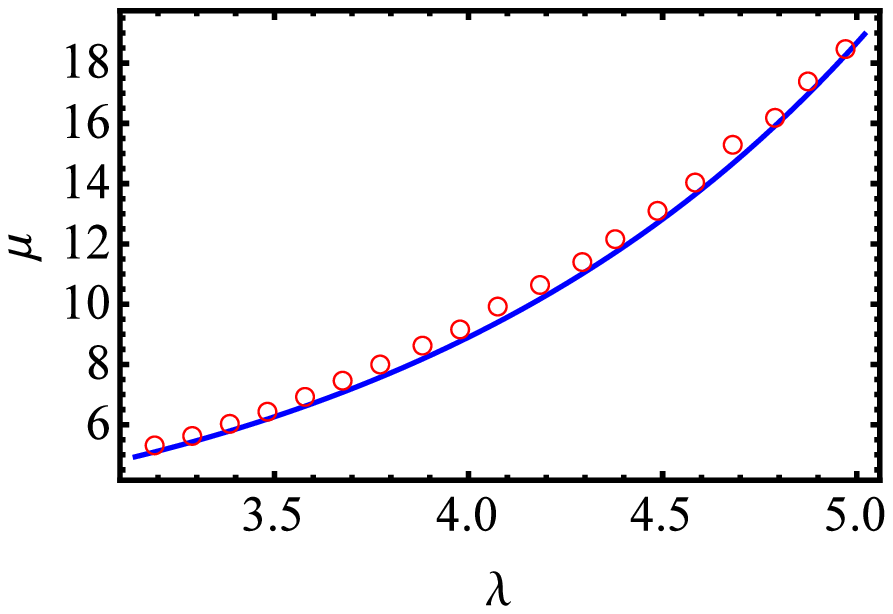}
        \label{comparison_mu}
    \end{subfigure}
    \quad\quad
    \begin{subfigure}{0.45\textwidth}
        \caption{}
        \includegraphics[width=\textwidth]{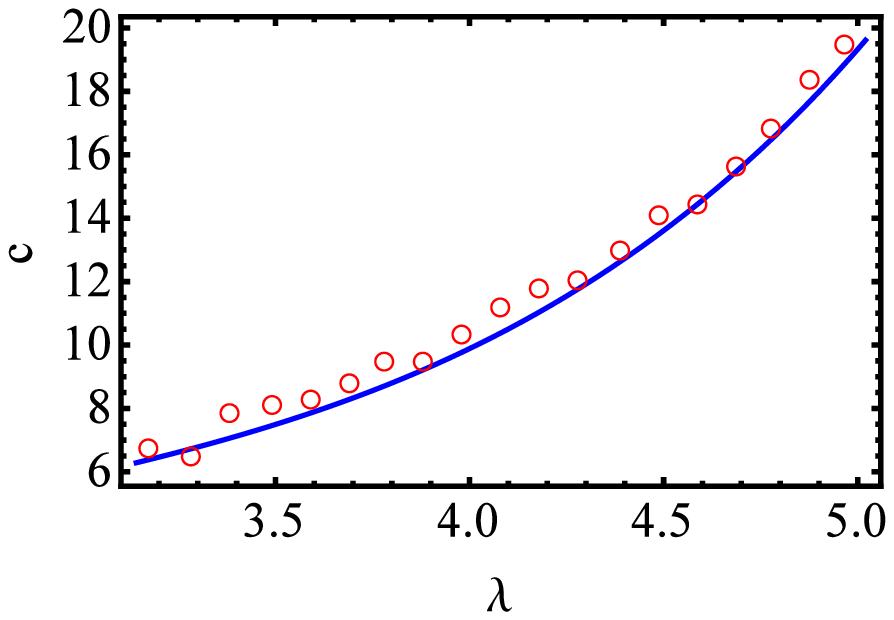}
        \label{comparison_c}
    \end{subfigure}
    \caption{The Legendre transform (\ref{mu_def}) of the action (a) and the TW velocity (b) versus $\lambda$. Solid line: our exact results. Circles: Monte-Carlo simulations \cite{Hurtado2011}.}
    \label{comparison_mu_c}
\end{figure}
\begin{figure}
    \centering
    \includegraphics[scale=1.0]{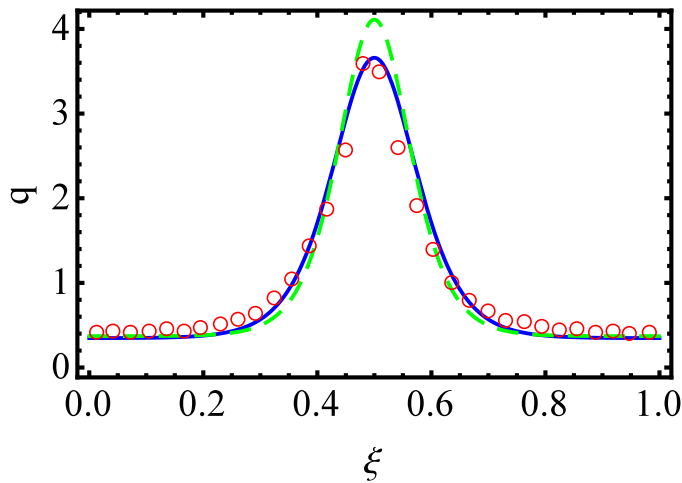}
    \caption{The TW density field $q(\xi)$. The solid and dashed lines show our exact solution and the large-current soliton approximation (\ref{ld_density_expansion}), respectively. The circles are the results of the Monte-Carlo simulations \cite{Hurtado2011}.  Here we recentered our density profiles to $\xi=0.5$.}
    \label{comparison_q}
\end{figure}
Hurtado and Garrido performed extensive Monte-Carlo simulations of the microscopic KMP model on a ring, employing a specialized algorithm  which amplifies rare large deviations of current \cite{Hurtado2011}. Their results are given in the form of the Legendre transform of our action:
\begin{equation}
    \label{mu_def}
    \mu(\lambda)=\max_{J}\left[ \lambda J - \frac{a}{T} S(J) \right] .
\end{equation}
We compared our analytic results with  their simulations with the maximum number of lattice sites, $N=32$, while setting $a=1/N$, $L=1$, $n_0=1$, and $D_0=1/2$. We
used Eq.~(\ref{mu_def}) to obtain the relation between $J$ and $\lambda$, and calculated our predictions for $\mu(\lambda)$ and $c(\lambda)$. (Note that, for these parameters, the critical value of $\lambda$ for the appearance of the TW is equal to $\pi$.) We then extracted the numerical results of Hurtado and Garrido for $\mu(\lambda)$ and $c(\lambda)$ in the supercritical region from Figs. 2 and 4 of  Ref.~\cite{Hurtado2011}. The resulting comparison is presented in Fig.~\ref{comparison_mu_c}, and a very good agreement is observed.

We also extracted the density profile $q(\xi)$, observed in a single Monte-Carlo realization for $\lambda=4.2$, from Fig. 3 of Ref.~\cite{Hurtado2011}, and compared it with our exact solution and to the large-current soliton approximation (\ref{ld_density_expansion}). Using the relation $J(\lambda)$, we found that $I(\lambda=4.2)\simeq 2.38I_c$. The comparison is presented in Fig.~\ref{comparison_q}. The exact solution shows a very good agreement with their results. Our analytical solution is also in perfect agreement with their numerical solution (not shown). The soliton approximation holds fairy well, in spite of the relatively low supercriticality.

\section{Discussion}
\label{disc}

In this work we investigated the statistics of large fluctuations of current in the one-dimensional Kipnis-Marchioro-Presutti (KMP)  model subject to periodic boundary conditions. We employed the macroscopic fluctuation theory (MFT) to derive the governing equations and boundary conditions for the optimal history of the system conditioned on a given current. We solved these equations analytically for arbitrary supercritical current, assuming a traveling wave (TW) solution.   We showed that the dynamical phase transition at $I=I_c$, as observed in the action $S(J)$, is of the second order.  We found simple asymptotics for the optimal history and for the action for weakly supercritical currents and for very large currents. The weakly super-crticial asymptotics are presented in Eqs.~(\ref{action_expansion}), (\ref{velocity_phase_expansion}) and (\ref{solution_density_expansion}); they extend  previous results of Bodineau and Derrida \cite{derrida2005} to higher orders in the supercriticality.

For very large currents,  the TW solution acquires the shape of a soliton, whereas the probability $P(J)$ behaves in the leading order as
$$
-\ln P (J) \simeq \frac{TJ}{a n_0}\, \ln \frac{L J}{D_0 n_0},
$$
as follows from Eqs.~(\ref{constraint_current_KMP_norm}), (\ref{action_TW}) and Eq.~(\ref{ld_extreme}). This result is strikingly similar
to the large-current asymptotic of the KMP model on an infinite line, when starting from a step-like initial condition \cite{meerson2013}.
Not surprisingly, the soliton-like solution of the MFT equations plays a crucial role in the latter problem too.

It would be interesting  to see whether the TW solution is indeed the true minimizer of the action at arbitrary supercriticality, which was the main assumption of this work. The Monte Carlo simulations of Hurtado and Garrido \cite{Hurtado2011} could only probe a limited range of supercritical currents. One way to proceed would be to numerically solve the complete MFT problem, formulated in Sec. \ref{MFT_general},  without making any assumption on the character of solution.

\section*{Acknowledgements}

We acknowledge a useful discussion with Pavel V. Sasorov. This research was supported by grant No. 2012145 from the United States-Israel Binational Science Foundation (BSF).

\appendix

\section{Derivation of the MFT equations}
\label{derivation}

Here we complete the derivation of equations (\ref{MFT_equation_q}) and (\ref{MFT_equation_v}), starting from Eq.~(\ref{probability_density}). In the saddle-point approximation, the probability of a specific current $J$ is equal to
\begin{equation}
    \label{probability_current}
    -\ln\left[P(J)\right] \, \simeq S(J)= \int_0^T \! dt \int_0^L \! dx \, \frac{\left[ j_{\text{opt}} + D(q_{\text{opt}})\partial_x q_{\text{opt}} \right]^2}{2\sigma(q_{\text{opt}})},
\end{equation}
where $q_{\text{opt}}(x,t)$ and $j_{\text{opt}}(x,t)$ are the optimal (i.e. most probable) histories of the fields $q$ and $j$ for a given $J$. We will drop the ``opt" subscript from now on.

To account for the connection between $q$ and $j$, as dictated by the continuity equation (\ref{continuity}), we introduce an auxiliary potential $\psi(x,t)$, defined by
\begin{equation}
    \label{psi_def}
              q = \partial_x \psi, \quad j = -\partial_t \psi.
\end{equation}
Now the action becomes
\begin{equation}
    \label{action_psi}
    S[\psi] = \int_0^T \! dt \int_0^L \! dx \, \frac{ \left[ D(\partial_x\psi)\partial_x^2\psi - \partial_t\psi \right]^2 }{2\sigma(\partial_x\psi)} = \int_0^T \! dt \int_0^L \! dx \, \frac{1}{2}\sigma(\partial_x\psi)\left(\partial_x p\right)^2,
\end{equation}
where we have defined the momentum density gradient
\begin{equation}
    \label{momentum_def}
    \partial_x p = \frac{ D(q)\partial_x q + j }{\sigma(q)} = \frac{ D(\partial_x\psi)\partial_x^2\psi - \partial_t \psi }{\sigma(\partial_x\psi)}.
\end{equation}
Now let us take a variation of the form $\psi(x,t) \rightarrow \psi(x,t) + \delta\psi(x,t)$ and evaluate the induced variation of the action to first order in $\delta\psi$:
\begin{align}
    \label{action_variation}
    \delta S = S[\psi+\delta\psi]-S[\psi] &= \int_0^T \! dt \int_0^L \! dx \, \left\{ -\frac{1}{2}\sigma'(\partial_x\psi)\left(\partial_x p \right)^2 \partial_x\delta\psi \right.\\
    \nonumber &\left.\vphantom{\frac{1}{2}} + \left(\partial_x p\right) \left[ D'(\partial_x\psi)\left(\partial_x^2\psi\right)\partial_x\delta\psi + D(\partial_x\psi)\partial_x^2\delta\psi - \partial_t\delta\psi \right] \right\}.
\end{align}
Before we integrate by parts to eliminate the derivatives of $\delta\psi$, we need to know what happens with the variation at the integration boundaries. To start with, $\partial_x\psi=q$ obeys the periodicity condition (\ref{boundary_cond}), and therefore also $\partial_x\delta\psi$, which results in $\partial_x\delta\psi(L,t) = \partial_x\delta\psi(0,t)$. Next, let us examine the energy conservation equation (\ref{constraint_mass}):
\begin{equation}
    \label{constraint_mass_psi}
    \frac{1}{L} \int_0^L \!dx \,\partial_x \psi = \frac{1}{L}\left[ \psi(L,t) - \psi(0,t) \right] = n_0.
\end{equation}
Taking the variation of it we obtain $\delta\psi(L,t) = \delta\psi(0,t)$. The variation at $t=0$ vanishes for the deterministic initial condition (\ref{initial_cond}). Now we integrate Eq.~(\ref{action_variation}) by parts and obtain
\begin{align}
    \label{action_variation_2}
    \delta S &= \int_0^T \! dt \int_0^L \! dx \, \left\{ \partial_x\left[ \frac{1}{2}\sigma'(\partial_x\psi)\left(\partial_x p \right)^2 \right] - \partial_x\left[ D'(\partial_x\psi)\left(\partial_x^2\psi\right)\partial_x p \right] \right. \\
    \nonumber &\left.\vphantom{\left[\frac{1}{2}\right]} + \partial_x^2\left[ D(\partial_x\psi)\partial_x p \right] + \partial_{xt}p \right\}\delta\psi - \int_0^L \! dx \,\left[\partial_x p(x,T)\right]\delta\psi(x,T).
\end{align}
The first MFT equation comes from spatial differentiation of Eq.~(\ref{momentum_def}). The second one comes from the demand that the double integral in (\ref{action_variation_2}) vanish for all $\delta\psi$. Then, the two MFT equations obtained are indeed Eqs. (\ref{MFT_equation_q}) and (\ref{MFT_equation_v}), where we defined $v = \partial_x p$. The current constraint (\ref{constraint_current}), which translates to
\begin{equation}
    \label{constraint_current_psi}
    - \frac{1}{LT} \int_0^T \!dt \int_0^L \!dx \,\partial_t \psi = - \frac{1}{LT} \int_0^L \!dx \,\left[ \psi(x,T) - \psi(x,0) \right] = J,
\end{equation}
can be imposed by introducing a Lagrange multiplier $\lambda$:
\begin{equation}
    \label{action_psi_2}
    S^*[\psi] = S[\psi] + \frac{\lambda}{LTJ} \int_0^L \!dx \,\left[ \psi(x,T) - \psi(x,0) \right],
\end{equation}
where $\lambda$ is defined as dimensionless. After performing the variation of $S^*$ and using $\delta\psi(x,0)=0$, this gives a contribution to the single integral in Eq.~(\ref{action_variation_2}). By demanding that it vanishes for all $\delta\psi$, the boundary condition (\ref{current_constraint_MFT}) for $v(x,T)$ is obtained.
Finally, the current density $j$ can be expressed through $v$ via the definition (\ref{momentum_def}), which results in Eq.~(\ref{current density}).

\section{TW solutions with shorter wavelengths}
\label{multiple_waves}

\begin{figure}[ht]
\includegraphics[scale=0.9]{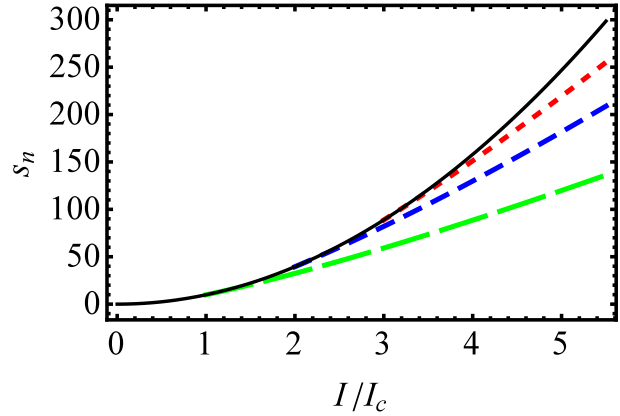}
\caption{The TW action $s_n$ versus $I/I_c$ for three different modes: the fundamental $n=1$ (long-dashed), $n=2$ (medium-dashed) and $n=3$ (short-dashed). The constant-density action is also shown, in solid. As one can see, for $I>I_c$ the action is minimal for $n=1$.}
\label{shorterTW}
\end{figure}

Let us return to the derivation in Sec. \ref{DODP}, but assume that the solution represents a mode with $n>1$ wavelengths, where $n$ is an integer. Now Eq.~(\ref{constraint_qs1}) becomes
\begin{equation}
    \label{constraint_qs1_n}
    \frac{2\text{K}(k)n}{\sqrt{C_2(q_3-q_1)}} = 1 .
\end{equation}
As one can easily check, Eq.~(\ref{constraint_qs2}) does not change. The changes to the solution (\ref{solution_density_final}) and the constraint (\ref{constraints10}) are then
\begin{equation}
    \label{solution_density_final_changes_n}
    q(\xi) \rightarrow q(n\xi), \quad v(\xi) \rightarrow nv(n\xi), \quad I \rightarrow \frac{I}{n}.
\end{equation}
Let us see how the action for the $n$-th mode, which we denote $s_n$, compares to  $s_1$:
\begin{equation}
    \label{action_changes_n}
    s_n(I) = \int_0^1 \! d\xi \, q^2(n\xi) n^2 v^2(n\xi) = \int_0^n \! \frac{d\bar{\xi}}{n} \, q^2(\bar{\xi}) n^2 v^2(\bar{\xi}) = n^2 \int_0^1 \! d\bar{\xi} \, q^2(\bar{\xi}) v^2(\bar{\xi}) = n^2 s_1\left(\frac{I}{n}\right).
\end{equation}
By virtue of Eqs.~(\ref{solution_density_final_changes_n}) and (\ref{action_changes_n}), the critical current for the $n$-th
mode is equal to $nI_c$. Therefore, at $I_c<I<2 I_c$  only the fundamental mode $n=1$ is possible. At $2 I_c<I<3 I_c$ there are two possible modes: $n=1$ and $2$. At $3 I_c<I<4 I_c$ there are three possible modes: $n=1$, $2$ and $3$, etc. 
For very large currents, $\ln I \gg 1$, we have $s_1(w) \simeq w \ln w$, see Eq.~(\ref{ld_extreme}). As $s_1(w)$ grows slower than quadratically with $w$, one has $s_n > s_1$ for $n>1$ in this limit. As we checked numerically, the inequality $s_n > s_1$ also holds at all supercritical currents. This is illustrated in Fig.~\ref{shorterTW} for $n=2$ and $3$.

\end{document}